\documentclass[useAMS,usenatbib,a4paper]{mn2e}

\usepackage{graphicx}
\usepackage{amssymb}
\usepackage{times}
\usepackage{color}

\newcommand{\cf}{\textit{cf.~}}
\newcommand{\ie}{\textit{i.e.,}~}
\newcommand{\eg}{\textit{e.g.,}~}
\newcommand{\ms}{{\rm ms}}

\newcommand{\Msun}{M_{\odot}}

\begin{document}

\title[A stability criterion for relativistic stars]
      {A quasi-radial stability criterion for rotating relativistic stars}

\author[K. Takami, L. Rezzolla and S. Yoshida]
{
Kentaro~Takami$^{1}$ 
Luciano Rezzolla$^{1,2}$ and
Shin'ichirou Yoshida$^{3}$
\\
$^{1}$Max-Planck-Institut f{\"u}r Gravitationsphysik, Albert Einstein Institut, Golm, Germany \\
$^{2}$Department of Physics, Louisiana State University, Baton Rouge, LA USA \\
$^{3}$Department of Earth Science and Astronomy, Graduate School of Arts and Sciences,
University of Tokyo, Japan}

\date{}

\pagerange{\pageref{firstpage}--\pageref{lastpage}} \pubyear{2011}

\maketitle

\label{firstpage}

\begin{abstract}
The stability properties of relativistic stars against gravitational
collapse to black hole is a classical problem in general relativity.
A sufficient criterion for secular instability was established by
Friedman, Ipser and Sorkin (1988), who proved that a sequence of
uniformly rotating barotropic stars is secularly unstable on one side
of a turning point and then argued that a stronger result should hold:
that the sequence should be stable on the opposite side, with the
turning point marking the onset of secular instability. We show here
that this expectation is not met. By computing in full general
relativity the $F$-mode frequency for a large number of rotating
stars, we show that the neutral-stability point, \ie where the
frequency becomes zero, differs from the turning point for rotating
stars. Using numerical simulations we validate that the new criterion
can be used to assess the dynamical stability of relativistic rotating
stars.

\end{abstract}


\section{INTRODUCTION}
\label{sec:INTRODUCTION}

The stability of a relativistic star against collapse to black hole is
one of the most important predictions of general relativity.  While
this problem is reasonably well understood for nonrotating
stars~\citep{Misner73}, this is not the case for rotating stars and is
particularly obscure when the stars are rapidly rotating.  A milestone
in this landscape is the criterion for secular stability proposed by
Friedman, Ipser and Sorkin (1988), who proved that a sequence of
uniformly rotating barotropic stars is secularly unstable on one side
of a turning point (an extremum of mass along a sequence of constant
angular momentum, or an extremum of angular momentum along a sequence
of constant rest-mass). They then argued, based on an expectation that
viscosity leads to uniform rotation, that the turning point should
identify the onset of secular instability.  While for nonrotating star
the turning point coincides with the secular-instability point (and
with the dynamical-instability point for a barotropic star if the
perturbation satisfies the same equation of state of the equilibrium
model), for rotating stars it is only a sufficient condition for a
secular instability. Lacking other guides, the turning point is
routinely used to find a dynamical instability in
simulations~\citep{Baiotti04, Radice:10}.

Our understanding of the dynamical instability of relativistic stars
in uniform rotation can be improved by determining the
neutral-stability line, that is the set of stellar models whose
frequency of the fundamental mode of quasi-radial oscillation
($F$ mode) is vanishingly small. While this problem is challenging
from a perturbative point of view, especially when the rate of
rotation becomes high, it can be tackled through numerical
calculations. We have therefore simulated in full general relativity
$54$ stellar models and calculated accurately the corresponding
$F$-mode frequency via a novel analysis of the power spectral density
(PSD) of the central rest-mass density. This new approach has been
validated through a comparison with all the available data, showing
excellent agreement and, most importantly, a much smaller variance. By
construction, in fact, simulations cannot evolve models at (or near)
the neutral-stability line, but the accuracy of our $F$-mode
frequencies and their smooth dependence on the central rest-mass
density and dimensionless rotation rate, have allowed us to produce an
analytic fit of the data and deduce from this the neutral-stability
line. We find in this way that it coincides with the turning point for
spherical stars, but not for rotating stars, with the difference
increasing with the angular momentum. Although somewhat surprising,
this difference is not in contrast with the predictions of the
turning-point criterion, since the latter is only a sufficient
condition for secular instability and not a necessary condition for
secular and dynamical instability.  Hence, a stellar model which is
stable according to the turning-point criterion, can be nevertheless
dynamical unstable.

To test the new stability line and validate whether it can be used to
mark the threshold for dynamical stability, we have evolved stellar
models whose properties fall in a small region near the two stability
lines. Special attention has been paid to stellar models that are
predicted to be stable by the turning-point criterion but unstable by
the neutral-stability line. Because these model indeed collapse to
black holes, we conclude that the neutral-stability line can be used
effectively to mark the boundary to dynamical instability.

The organization of the paper is as
follows. Section~\ref{sec:METHOD_AND_MODEL} describes the numerical
setup and initial data, while Sect.~\ref{sec:method_accuracy} presents
our approach to extract the eigenfrequency and offers comparisons with
previous work. Section~\ref{sec:RESULTS} collects our results and a
comparison between the two stability criteria, leaving the conclusions
to Sect.~\ref{sec:SUMMARY}. Unless stated differently, we use units in
which $c=G=M_{\sun}=1$.
%

\section{Numerical Setup and Initial Data}
\label{sec:METHOD_AND_MODEL}

All of our calculations have been performed in full general relativity
(GR) using the \texttt{Whisky2D} code described in detail
in~\citet{Kellermann:08a}. This is a 2-dimensional (2D) code based on
the 3-dimensional (3D) \texttt{Whisky} code~\citep{Baiotti04}, and
exploiting the condition of axisymmetry through the ``cartoon''
method~\citep{Alcubierre01b}. In essence, the evolution of the
spacetime is obtained using the 2D version of \texttt{Ccatie}, a
finite-differencing code providing the solution of a conformal
traceless formulation of the Einstein
equations~\citep{Pollney:2007ss}, while the equations of relativistic
hydrodynamics are solved a flux-conservative formulation of the
equations, as first discussed in detail in~\citet{Baiotti04}. The {\tt
  Whisky2D} code implements a variety of approximate Riemann solvers
and several reconstruction methods and, as discussed
in~\cite{Giacomazzo:2009mp}, the use of reconstruction schemes of
order high enough is fundamental for an accurate evolution.  In
particular, the results presented here have been computed using the
piecewise-parabolic reconstruction method(PPM)~\citep{Colella84}, the
HLLE approximate Riemann solver~\citep{Harten83}, and a 3rd-order
Runge-Kutta method for the time evolution.

The initial equilibrium stellar models are built using the
\texttt{rns} code~\citep{Stergioulas95} as isentropic, uniformly
rotating relativistic perfect-fluid polytropes with equation of state
\begin{equation}
p = K \rho ^{\Gamma}\,,\hskip 2.0cm e=\rho+\frac{p}{\Gamma-1}\,,
\end{equation}
where $p$ is the pressure, $\rho$ the rest-mass density, $K$ the
polytropic constant, $\Gamma$ the polytropic exponent, and $e$ the
energy density. Although all the results can be rescaled for any
choice of $K$ and $\Gamma$, we have here set $K=100$ and $\Gamma=2$,
which yield stars with maximum gravitational mass is $M =
1.64\,M_{\odot}$ for a nonrotating star and $M = 1.88\,M_{\odot}$ for
a uniformly rotating one. The \texttt{rns} code provides an
equilibrium solution in spherical polar coordinates after specifying
for each stellar model a central density $\rho_{c}$ and a equatorial
and polar (coordinate) radii in a ratio $r_p/r_e$. Once this solution
is found, it is mapped to a Cartesian grid of \texttt{Whisky2D} and
used as initial data for the subsequent evolution. Attention needs to
be paid that the resolution in the calculation of the initial data
matches well the one used in the evolution. We have verified that a
resolution of $(n_{r}, n_{\theta})= (2001,2601)$ [$(n_{r},
  n_{\theta})= (1001,1301)$], with $(n_{r}, n_{\theta})$ the number of
points of the radial and angular grids of the \texttt{rns} code, are
needed for an accurate evolution in the high [low]-resolution setup of
the \texttt{Whisky2D} code. Furthermore, because we are not interested
here in extracting gravitational-wave information, we place the outer
boundary at a few stellar radii and use a uniform grid with spacing
$\Delta x = \Delta z = h$ ranging between $h = 0.04\,\Msun$ for the
rapidly rotating models and up to $h= 0.1\,\Msun$ for the slowly
rotating ones. As done in~\citet{Kellermann:08a}, we stagger the grid
in the $x$-direction of half a cell. A large number of tests have been
carried out to verify that the results do not depend on the position
of the outer boundary, or on the value of the density in the
atmosphere (see~\citet{Baiotti04}), which we set to be $9$ orders of
magnitude smaller than the central one.

As discussed by many authors~\citep{Font99,Font02c,Baiotti04}, the
truncation error in the initial data is sufficient to trigger
perturbations in the star, which will start to oscillate in a number
of eigenmodes. However, because we need to determine the
eigenfrequency of the $F$ mode, it is important that as much as
possible of the initial perturbation energy goes into exciting that
mode. For this reason we introduce an initial perturbation using
the eigenfunction of the $F$ mode for a nonrotating neutron star with
the same central density, and which can be computed from linear
perturbation theory. More specifically, denoting with
$\psi_{_{\mathrm{TOV}}}$ any fluid quantity of the nonrotating model
with the same central density and with $\delta
\psi_{_{\mathrm{TOV}}}(r)$ the corresponding eigenfunction with $r$
the radial coordinate in isotropic coordinate system, we as
approximate equivalent eigenfunction for a rotating star in a
coordinate system $(r, \theta)$ as $\delta \psi(r,\theta)= \delta
\psi_{_{\mathrm{TOV}}}(r R_{_{\mathrm{TOV}}}/R(\theta))$, where
$R_{_{\mathrm{TOV}}}$ is the radius of the nonrotating star and
$R(\theta)$ that of the rotating star, which will obviously depend on
the angle $\theta$. As a result, the power in the initial perturbation
is mostly concentrated in the $F$ mode, whose corresponding peak in
the PSD of any hydrodynamical quantity is larger by at least a factor
$10$ than any other mode. As an additional validation of the
procedure, we have computed the numerical eigenfunction for some
selected models and verified that it matches very well the guessed one
even in the case of rapidly rotating stars and long-term evolutions.

\section{Methodology and Accuracy}
\label{sec:method_accuracy}

As customary, we extract the $F$-mode frequency by performing a
discrete Fourier transform of the evolution of a representative
hydrodynamical quantity, such as the central rest-mass density
$\rho_{\mathrm{c}}$, and by inspecting the corresponding PSD. Defining
as $F_\mathrm{N}$ the frequency of the largest peak in the numerical
PSD, previous studies determined the value of the $F$-mode frequency,
$F$, by fitting the PSD with a known analytic function [\eg a
  Lorentzian,~\citep{Kellermann:08a}] or by taking the derivative of
the PSD~\citep{Zink:2010a}. The frequency obtained in these ways
depends sensitively on the fitting function used, on the shape of the
PSD around $F_\mathrm{N}$, and on the evolution time $\tau$. We here
use a different approach. Because $F_\mathrm{N}$ will tend to $F$ as
the evolution time $\tau \rightarrow \infty$, we simply consider the
evolution of $F_\mathrm{N}$ for increasingly large values of
$\tau$. What we find in this way is that $F_\mathrm{N}(\tau)$ is an
oscillating function around $F$, whose amplitude is however bounded by
two envelopes which have a clear $1/\tau$ dependence. Fitting for
these envelopes and extrapolating for $\tau \rightarrow \infty$ we
obtain a very accurate and possibly optimal value for $F$. As we will
discuss in the following Section, this approach turns out to give an
excellent measure of the $F$-mode eigenfrequency and we recommend it
in all those studies aimed at determining eigenfrequencies of
relativistic stars.

\subsection{Comparison with previous works}

The $F$-mode frequency of spherical stars can be computed to arbitrary
precision within a linear perturbative approach (see~\citet{Yoshida01}
and references therein). Hence, as a first validation of the accuracy
of our procedure we have estimated the $F$-mode frequency for $14$
nonrotating models with $\rho_{c}\in[3.0\times 10^{-4}, 3.0\times
  10^{-3}$]; in this range the $F$ mode first grows, then reaches a
maximum, and finally decreases to zero at the secular-instability
point, around $\rho_{c} \thickapprox 3.18 \times 10^{-3}$. Defining
the relative error as $\sigma_\mathrm{rel} \equiv
[(F)^2_{_{\mathrm{PT}}} - (F)^2_{_{\mathrm{N}}}] /
(F)^2_{_{\mathrm{PT}}}$, where $F_{_{\mathrm{PT}}}$ and
$F_{_{\mathrm{N}}}$ are respectively the frequencies of the $F$ mode
from perturbation theory and from our simulation.  The relative error
is extremely small at low densities (\eg $\sigma_\mathrm{rel} \lesssim
0.005$ for $\rho_{c} \thickapprox 0.3 - 1.5 \times 10^{-3}$) and it
increases with the density (\eg $\sigma_\mathrm{rel} \lesssim 0.05$
for $\rho_{c} \thickapprox 2.5 - 3.0 \times 10^{-3}$), becoming of the
order of about $10\%$ at the edge of the secular instability. This is
obviously due to the fact that as $F_\mathrm{N} \thickapprox 0$,
numerical calculations become increasingly long and inaccurate.

\begin{figure}
 \includegraphics[width=7.5cm]{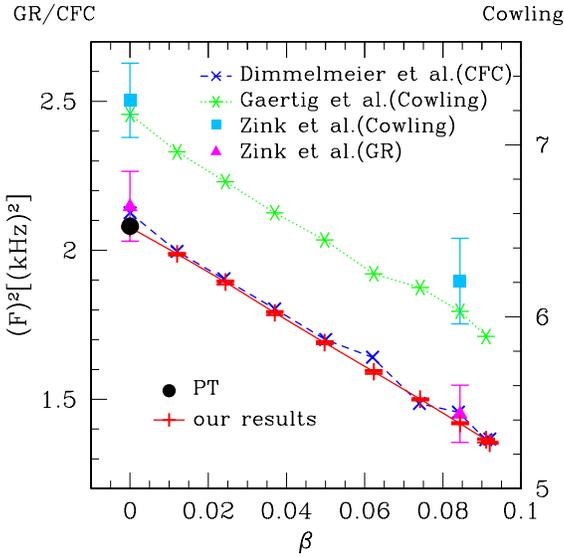}
 \caption{Comparison of our $F$-mode frequencies with those of
   previous works in either perturbation theory (PT), the CFC
   approximation~\citep{Dimmelmeier06}, the Cowling
   approximation~\citep{Gaertig:2008a,Zink:2010a}, or in full
   GR~\citep{Zink:2010a}.
 \label{fig:Comparison___Full_vs_CFC_vs_Cowling}}
\end{figure}

%
We next compare our numerical estimates for the $F$-mode frequency
with those made in several different approaches and approximations,
using as reference a central rest-mass density
\mbox{$\rho_{c}=1.28\times10^{-3}$}, as this is the one most commonly
used. We start our comparison by considering the case of nonrotating
stars, for which results are available from works
of~\citet{Dimmelmeier06} in either the conformally-flat condition
(CFC), or of~\citet{Zink:2010a} in full GR. This is shown in
Fig.~\ref{fig:Comparison___Full_vs_CFC_vs_Cowling}, which reports the
$F$-mode frequency as a function of the dimensionless ratio $\beta
\equiv T/|W|$ between the rotational kinetic energy $T$ and the
binding energy $W$. Note that the frequency is reported in two
different scales, referring to simulations either in full GR/CFC (left
scale) or in the Cowling approximation (right scale), which
systematically yields larger frequencies. Although the CFC (blue
crosses) for a nonrotating should give the same frequency in full GR
(magenta filled triangles and red crosses) and in perturbation theory
(black filled circle),
Fig.~\ref{fig:Comparison___Full_vs_CFC_vs_Cowling} shows that this is
not quite the case, although the differences are only of $\sim
2\%$. Considerably larger are instead the differences with the
frequencies in the Cowling approximation, which are larger of a factor
of $\sim 3$ (green stars and light-blued filled squares). Clearly, the
difference between the results in full GR and the perturbative ones is
much smaller and indeed the one with our new results is the smallest
among all the data available. We also note that our results also
report the estimated error bars, which are much smaller than the size
of the symbols.

Considering next the comparison also for rotating stars, it is easy to
see that our results in two dimensions match well those 
in three dimensions of~\citet{Zink:2010a}
for the rotation rates available and obviously have smaller error
bars. The very good match with the results in the
CFC~\citep{Dimmelmeier06}, with differences of a few percent only for
all the values of $\beta$, confirms the conclusions drawn
by~\citet{Dimmelmeier06}, that the CFC is a very good approximation,
at least for the dynamics of isolated
stars. Figure~\ref{fig:Comparison___Full_vs_CFC_vs_Cowling} also shows
that the comparison with frequencies computed in the Cowling
approximation~\citep{Gaertig:2008a,Zink:2010a} is considerably worse.
Besides an intrinsic difference between the two sets of data (the
frequencies of~\citet{Gaertig:2008a} are agreement only within the
error bars of~\citet{Zink:2010a}), the rate of change of the
frequencies with $\beta$ differs from the one found in full GR, being
less rapid for the latter (this is not evident because the figure has
two different vertical scales).  This comparison shows the Cowling
approximation to be inaccurate for all rotation rates.

In summary, this comparison validates our approach, highlighting its
accuracy and smoothness when compared to alternative methods. This
will be essential to find the neutral-stability line.

\section{RESULTS}
\label{sec:RESULTS}

As mentioned above, the space of parameters is spanned by central
rest-mass density and the angular momentum of the rotating models. To
cover the largest possible region of parameters we have evolved $54$
stellar models of relativistic stars with $\rho_{c}$ in the
range\footnote{Note that $\rho_{c}=1.0\times 10^{-3} \simeq 0.62\times
  10^{15}\,{\rm g/cm}^3$ and that $\rho_{\mathrm{max}}$ also marks the
  secular stability point for a nonrotating star.}
$[\rho_{\mathrm{min}}, \rho_{\mathrm{max}}]= [8\times 10^{-4},
  3.18\times 10^{-3}]$ and dimensionless rotation parameter $\beta$
between zero and the mass-shedding limit for the corresponding
sequence of constant central rest-mass density ($\beta=0.095$ is the
largest value considered). In this way we computed stellar models with
masses in the range $M/M_{\odot} \in [1.1,1.9]$.

\begin{figure}
 \includegraphics[scale=0.85]{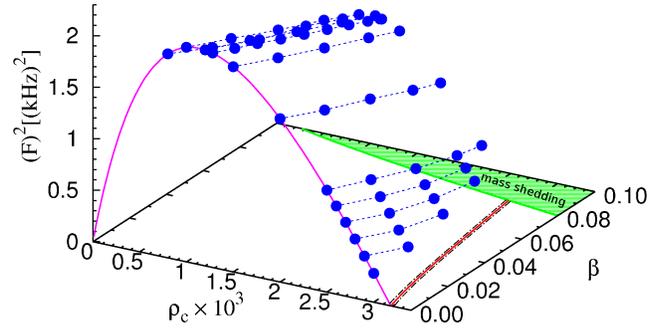}
 \caption{Square of the $F$-mode frequencies (blue filled circles) as
   a function of $\rho_{c}$ and $\beta$. The dashed green area shows
   models above the mass-shedding limit, and the red solid line
   marks the neutral stability (\cf Fig.~\ref{fig:Stable_Line}).
 \label{fig:F_surface2D3D_UniRot_PolyK100G2}}
\end{figure}

%
We show as filled blue circles in
Fig.~\ref{fig:F_surface2D3D_UniRot_PolyK100G2} all of the computed
$F$-mode frequencies, where the squares of the $F$-mode frequencies
$(F)^2$ are reported as function of $\rho_{c}$ and $\beta$. Shown as a
solid magenta line is the analytic fitting of the frequency for
nonrotating stars, while dashed blue lines show sequences of rotating
stars having the same rest-mass density. All models simulated have
nonzero $F$-mode frequencies and their number diminishes for
$(F)^2\approx 0$. As mentioned above, this is because for these models
the oscillation timescale tends to become extremely large (diverging
for $F=0$), thus becoming intractable in numerical simulations. In
addition, models near the neutral point could also be artificially
induced to collapse simply by the accumulation of the truncation error
(see also~\citet{Shibata:2003iy}), thus preventing any reliable
measure. As a result, our analysis has been constrained to values of
the frequencies $F \gtrsim 2.2 \times 10^{-3} \simeq 0.45\,{\rm kHz}$.
Fortunately, however, the quality of the data and the smoothness in
which they appear in Fig.~\ref{fig:F_surface2D3D_UniRot_PolyK100G2},
allow us to compute an analytic fit of the function
$(F)^2=(F)^2(\rho_{c}, \beta)$ and thus determine analytically the
neutral-stability line where $(F)^2=0$.

It is convenient to use a fitting function
$(F_\mathrm{fit})^2(\rho_{c},\beta)$ that is linear in $\beta$ and
such that \mbox{$(F)^2_\mathrm{fit}(\rho_{\mathrm{max}},0) = 0$} by
construction
\begin{eqnarray}
(F)^2_\mathrm{fit}(\rho_{c},\beta) &=& (F)^2_\mathrm{\mathrm{fit}}(\rho_{c},0) +
\beta \sum_{n=0}^5 b_{n} (\rho_{c})^n \\
& = &\sum_{n=0}^5 a_n (\rho_{c})^n + \beta \sum_{n=0}^5 b_{n} (\rho_{c})^n\,,
\label{eq:empirical}
\end{eqnarray}
where $a_n, b_n$ are constant coefficients, which a least-square
fitting with the data reveals to be
\begin{eqnarray*}
 && \hskip -0.65cm
    a_5 =  6.978 \times 10^8    \,, ~
    a_4 = -7.757 \times 10^6    \,, ~
    a_3 =  3.621 \times 10^4    \,, \\
 && \hskip -0.65cm
    a_2 = -9.599 \times 10      \,, ~
    a_1 =  1.172 \times 10^{-1} \,,  ~
    a_0 =  2.110 \times 10^{-7} \,,  \\
 && \hskip -0.65cm
  b_{5} = -5.599 \times 10^{10} \,, ~
  b_{4} =  4.862 \times 10^{8}  \,, ~
  b_{3} = -1.612 \times 10^{6}  \,, \\
 && \hskip -0.65cm
  b_{2} =  2.545 \times 10^{3}  \,, ~
  b_{1} = -1.896                \,, ~
  b_{0} =  3.357 \times 10^{-4} \,. 
\end{eqnarray*}

A confirmation of the accuracy of the ansatz (\ref{eq:empirical})
comes from the very small variance of a comparison with perturbative
results for nonrotating stars. Considering in fact over $90$ stellar
models with $\rho_\mathrm{c} \in [1.0\times 10^{-5}$, $3.182 \times
  10^{-3}]$, we obtain $\sigma_\mathrm{fit} \equiv \left|
(F)^2_{_{\mathrm{PT}}} - (F)^2_{_{\mathrm{fit}}}(\rho_{c},0) \right|
\lesssim 2 \times 10^{-7} \simeq 8\times 10^{-3}\,({\rm
  kHz})^2$. Similarly, when comparing over the whole set of numerical
data we find a variance that, as expected, is greater for large values
of $\rho_{c}$ and $\beta$ but that, overall, is $\sigma_\mathrm{fit}
\lesssim \sigma_\mathrm{max} \thickapprox 1 \times 10^{-6}$. Note that
these errors are smaller or at most comparable with the numerical
error bar, highlighting the quality of the fit.

\begin{figure}
 \includegraphics[width=7.5cm]{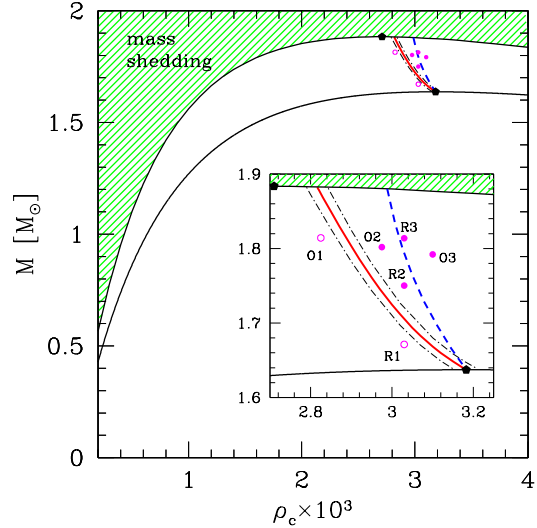}
 \caption{Stability lines in a $(\rho_{c},M)$ diagram. The two solid
   black lines mark sequences with either zero (lower line) or
   mass-shedding angular momentum (upper line), with the filled
   symbols marking the corresponding maximum masses. The solid red
   line is the neutral-stability line, ``thickened'' by the error bar
   (black dot-dashed lines). The blue dashed line is instead the
   turning-point criterion for secular stability. Marked with empty or
   filled circles are representative models with constant angular
   velocity \texttt{O1}, \texttt{O2}, \texttt{O3}, or constant initial
   central rest-mass density \texttt{R1}, \texttt{R2}, \texttt{R3}.
 \label{fig:Stable_Line}}
\end{figure}

Using expression (\ref{eq:empirical}), it is straightforward to
compute the neutral-stability line in a $(\rho_{c},\beta)$ plane as
the one at which $(F)^2_\mathrm{fit}(\rho_{c},\beta)=0$. Of course
this line will be ``thickened'' by the uncertainty associated to the
fit which, to be conservative, we consider to be
$\sigma_{\mathrm{max}}$ (We note that the thickness is much smaller
for $\beta \approx 0$ but it may be larger at high $\beta$ as a result
of the extrapolation.). While a neutral-stability line is already very
informative in a $(\rho_{c},\beta)$ plane, its greatest impact can be
appreciated in the more traditional $(\rho_{c},M)$ diagram. This is
shown in Fig.~\ref{fig:Stable_Line}, where the two solid black lines
refer to sequences of nonrotating (lower line) and mass-shedding
models (upper line), respectively. Drawn as solid red is the
neutral-stability line ``thickened'' by the error bar
$\sigma_{\mathrm{max}}$ (black dot-dashed lines). Finally, shown as a
blue dashed line is the turning-point criterion for secular stability
along a sequence of constant angular momentum $J$, \ie
$\left({\partial M}/{\partial \rho_{c}}\right)_{J=\mathrm{const}}=0$.

Clearly, the new neutral-stability criterion does coincide with the
turning-point criterion for nonrotating stars (\cf small inset), but
it differs from it as the angular momentum is increased, moving to
smaller central rest-mass densities. While unexpected, this difference
does not point to a conflict between the two criteria. This is because
the turning-point criterion is only a \textit{sufficient} condition
for secular instability of rotating stars; stated differently, while a
rotating stellar model which is at or to the right of the
turning-point line is expected to be also secular unstable, the
opposite is not true. Hence, the two criteria are compatible as long
as the secular instability line lies to the left (\ie for smaller
central rest-mass densities) of the neutral-stability
line. Determining the secular-stability line requires to consider a
dissipative mechanism such as viscosity, which is however absent in
our perfect-fluid description and difficult to introduce within a
fully relativistic hyperbolic description. However, because a
dynamically unstable model should also be secularly unstable, we in
fact expect the secular stability line to coincide or to be on the
left of the neutral-stability line. In other words, along a $J={\rm
  const.}$ sequence of stellar models we expect the following order
with increasing rest-mass density: secular instability, dynamical
instability, turning-point.

\begin{figure}
 \includegraphics[width=7.5cm]{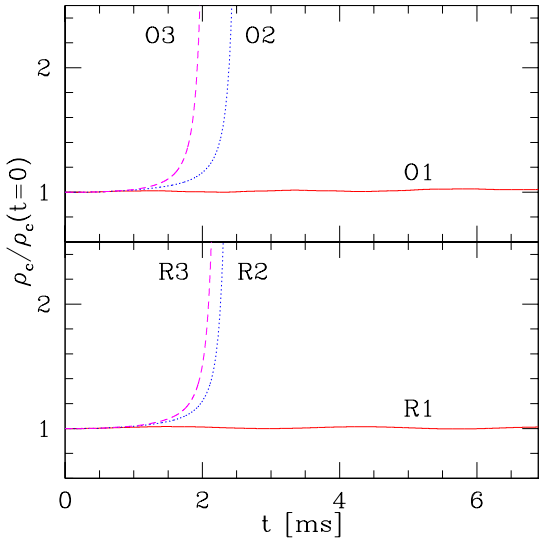}
 \caption{Evolution of $\rho_{\mathrm{c}}$ for models with constant
   angular velocity (upper panel) or constant initial central
   rest-mass density (lower panel). An exponential growth signals the
   collapse to black hole (\cf Fig. \ref{fig:Stable_Line}).
 \label{fig:Stable_Line_Confirm}}
\end{figure}

To validate that the neutral-stability line should be used in place of
the turning-point line to distinguish stellar models that are
dynamically unstable from those that are instead stable, we have
considered $6$ representative models whose properties fall in a small
region near the two stability lines. More specifically, we consider
two different sequences having either constant angular velocity, \ie
models \texttt{O1}, \texttt{O2}, \texttt{O3} in
Fig.~\ref{fig:Stable_Line_Confirm}, or constant $\rho_{c}$, \ie models
\texttt{R1}, \texttt{R2}, \texttt{R3}. The predictions for these
models are different according to which criterion is used for
stability. In fact, while models \texttt{O1}, \texttt{R1} are expected
to be stable for both criteria and models \texttt{O3}, \texttt{R3} are
expected to be unstable for both criteria, models \texttt{O2},
\texttt{R2} are predicted to be stable on a dynamical timescale by the
turning-point criterion but unstable by the neutral-stability
criterion.

To test these predictions we have evolved these configurations
maintaining the same computational setup (but without an initial
perturbation) and collected the corresponding evolution of the central
rest-mass density in Fig.~\ref{fig:Stable_Line}. As expected, models
\texttt{O1}, \texttt{R1} are found to be stable over about $7\,\ms$ as
indicated by the central rest-mass density that remains constant
(modulo the $F$-mode oscillations), while models \texttt{O3},
\texttt{R3} are found to collapse to black holes in less than $2\,\ms$
as indicated by the exponential increase of the rest-mass density (see
also~\cite{Baiotti04, Radice:10}). Similarly, models \texttt{O2},
\texttt{R2} are also found to collapse to black holes over a timescale
which is only slightly larger than that of models \texttt{O3},
\texttt{R3}. After validating that these results do not depend on the
specific numerical setup used (\eg placement of outer boundaries,
resolution or density in the atmosphere), we conclude that the
neutral-stability line can indeed be used to mark the boundary of a
dynamically unstable region.

\section{Conclusions}
\label{sec:SUMMARY}

The stability of rotating relativistic stars against gravitational
collapse to black hole is an old problem in general relativity,
impacting all those astrophysical problems where a neutron star may be
produced and induced to collapse as a result of mass
accretion. Despite the importance of this problem, no analytic a
criterion is known for a dynamical stability of rotating
stars. Important progress was made about $20$ years ago, when a
criterion for secular stability was proposed by~\citet{Friedman88},
who suggested that a turning point along a sequence of stellar models
with constant angular momentum can be associated with the onset of a
secular instability. Although this criterion is only a sufficient
condition for the development of a secular instability, it has been
systematically used to limit the region of dynamical instability
in simulations of relativistic stars~\citep{Baiotti04, Radice:10}.

To improve our understanding of the dynamical instability of
relativistic stars in uniform rotation, we have computed the
neutral-stability point for a large class of stellar models, \ie the
set of stellar models whose $F$-mode frequency is vanishingly small
(in a nonrotating star this point marks the dynamical stability
limit). More specifically, we have evolved in full general relativity 
$54$ stellar models and calculated the corresponding
$F$-mode frequency via a novel analysis of the PSD of the central
rest-mass density. Although our simulations cannot evolve models near
the neutral-stability line, the high accuracy of our estimates for the
eigenfrequencies (which have been validated through a comparison with
all the available data) and their regular dependence on the central
rest-mass density and dimensionless rotation rate, have allowed us to
produce an analytic fit of the data and deduce from this the
neutral-stability line. The latter coincides with the turning-point
line of~\citet{Friedman88} for nonrotating stars, but differs from it
as the angular momentum is increased, being located at smaller central
rest-mass densities as the angular momentum is increased. This
difference does not contradict turning-point criterion since the
latter is only a sufficient condition for secular instability.

To test this result we have evolved stellar models whose properties
fall in a small region near the two stability lines, paying special
attention to those stellar models that are predicted to be stable on a
dynamical timescale by the turning-point criterion but unstable by the
neutral-stability line. Numerical evidence that these model do
collapse to black holes allows us to conclude that the
neutral-stability line can be used effectively to mark the boundary to
dynamical instability. Besides improving our understanding of the
stability of relativistic stars, these results show that producing
black holes via the gravitational collapse of a neutron star is
simpler than expected. Furthermore, they can serve as a guide when
determining the neutral-stability line via perturbative techniques or
when extending it to differentially rotating stars.

\section*{Acknowledgments}

%
We are grateful to John Friedman and Nikolaos Stergioulas for extended
discussions and useful suggestions that have improved the
manuscript. We also thank Thorsten Kellermann for his work on the {\tt
  Whisky2D} code. Support comes also from the DFG grant
SFB/Transregio~7 and by ``CompStar'', a Research Networking Programme
of the European Science Foundation. KT is supported by a JSPS
Postdoctoral Fellowship for Research Abroad.
%

\bibliographystyle{mn2e}
%


\label{lastpage}

\end{document}